\documentclass[aps,prl,showpacs,twocolumn,groupedaddress]{revtex4}  % for submission
\usepackage{graphicx}% Include figure files
\usepackage{dcolumn}% Align table columns on decimal point
\usepackage{bm}% bold math
\usepackage{epsfig}

\begin{document}
%\linenumbers

%
%      %%%%%%%%%%%%%%%%%%%%%%%%%%%%%%%%%%%
%%%%%%%%%%%% SOME USEFUL LaTeX2e DEFINITIONS %%%%%%%%%%%%%%
%      %%%%%%%%%%%%%%%%%%%%%%%%%%%%%%%%%%%
%
\newcommand{\comm}[1]{\mbox{\mbox{\textup{#1}}}}
\newcommand{\subs}[1]{\mbox{\scriptstyle \mathit{#1}}}
\newcommand{\subss}[1]{\mbox{\scriptscriptstyle \mathit{#1}}}
\newcommand{\Frac}[2]{\mbox{\frac{\displaystyle{#1}}{\displaystyle{#2}}}}
\newcommand{\LS}[1]{\mbox{_{\scriptstyle \mathit{#1}}}}
\newcommand{\US}[1]{\mbox{^{\scriptstyle \mathit{#1}}}}
\def\gsim{\mathrel{\rlap{\raise.4ex\hbox{$>$}} {\lower.6ex\hbox{$\sim$}}}}
\def\lsim{\mathrel{\rlap{\raise.4ex\hbox{$<$}} {\lower.6ex\hbox{$\sim$}}}}
\renewcommand{\arraystretch}{1.3}
\newcommand{\Edep}{\mbox{E_{\mathit{dep}}}}
\newcommand{\Ebeam}{\mbox{E_{\mathit{beam}}}}
\newcommand{\Exrc}{\mbox{E_{\mathit{rec}}^{\mathit{exp}}}}
\newcommand{\Emrc}{\mbox{E_{\mathit{rec}}^{\mathit{sim}}}}
\newcommand{\Evis}{\mbox{E_{\mathit{vis}}}}
\newcommand{\Edepi}{\mbox{E_{\mathit{dep,i}}}}
\newcommand{\Evisi}{\mbox{E_{\mathit{vis,i}}}}
\newcommand{\Exrci}{\mbox{E_{\mathit{rec,i}}^{\mathit{exp}}}}
\newcommand{\Emrci}{\mbox{E_{\mathit{rec,i}}^{\mathit{sim}}}}
\newcommand{\Etmis}{\mbox{E_{\mathit{t,miss}}}}
%
%%% D0 and the physics analyses related stuff:
%
\newcommand{\lt}{\mbox{$<$}}
\newcommand{\gt}{\mbox{$>$}}
\newcommand{\lte}{\mbox{$\le$}}
\newcommand{\gte}{\mbox{$\ge$}}

\newcommand{\xa}{\mbox{$x_{a}$}}
\newcommand{\xb}{\mbox{$x_{b}$}}
\newcommand{\xp}{\mbox{$x_{p}$}}
\newcommand{\xpb}{\mbox{$x_{\bar{p}}$}}

\newcommand{\alphas}{\mbox{$\alpha_{s}$}}
\newcommand{\partt}{\mbox{$\partial^{t}$}}
\newcommand{\partd}{\mbox{$\partial_{t}$}}
\newcommand{\parmut}{\mbox{$\partial^{\mu}$}}
\newcommand{\parmud}{\mbox{$\partial_{\mu}$}}
\newcommand{\parnut}{\mbox{$\partial^{\nu}$}}
\newcommand{\parnud}{\mbox{$\partial_{\nu}$}}
\newcommand{\Amut}{\mbox{$A^{\mu}$}}
\newcommand{\Amud}{\mbox{$A_{\mu}$}}
\newcommand{\Anut}{\mbox{$A^{\nu}$}}
\newcommand{\Anud}{\mbox{$A_{\nu}$}}
\newcommand{\Fmunut}{\mbox{$F^{\mu\nu}$}}
\newcommand{\Fmunud}{\mbox{$F_{\mu\nu}$}}
\newcommand{\Gmut}{\mbox{$\gamma^{\mu}$}}
\newcommand{\Gmud}{\mbox{$\gamma_{\mu}$}}
\newcommand{\Gnut}{\mbox{$\gamma^{\nu}$}}
\newcommand{\Gnud}{\mbox{$\gamma_{\nu}$}}
\newcommand{\pvecsq}{\mbox{$\vec{p}^{\:2}$}}
\newcommand{\bigsum}{\mbox{$\displaystyle{\sum}$}}
%
% Triggers:
\newcommand{\Lzero}{Level \O}
\newcommand{\Lone}{Level $1$}
\newcommand{\Ltwo}{Level $2$}
\newcommand{\Lhalf}{Level $1.5$}
%
% D0 General
\newcommand{\bzero}{\mbox{\comm{B\O}}}
\newcommand{\dzero}{\mbox{D\O}}
\newcommand{\dzerosm}{\mbox{\comm{$\scriptsize{D\O}$}}}
\newcommand{\runb}{Run~$1$B}
\newcommand{\runa}{Run~$1$A}
\newcommand{\runone}{Run~$1$}
\newcommand{\runtwo}{Run~$2$}
\newcommand{\dzpjet}{\textsc{D{\O}Pjet}}
\newcommand{\y}{\mbox{$y$}}
\newcommand{\z}{\mbox{$z$}}
\newcommand{\px}{\mbox{$p_{x}$}}
\newcommand{\py}{\mbox{$p_{y}$}}
\newcommand{\pz}{\mbox{$p_{z}$}}
\newcommand{\ex}{\mbox{$E_{x}$}}
\newcommand{\ey}{\mbox{$E_{y}$}}
\newcommand{\ez}{\mbox{$E_{z}$}}
\newcommand{\et}{\mbox{$E_{T}$}}
\newcommand{\etprime}{\mbox{$E_{T}^{\prime}$}}
\newcommand{\etone}{\mbox{$E_{T}^{\mathrm{1}}$}}
\newcommand{\ettwo}{\mbox{$E_{T}^{\mathrm{2}}$}}
\newcommand{\etlj}{\mbox{$E_{T}^{\subs{lj}}$}}
\newcommand{\etmax}{\mbox{$E_{T}^{max}$}}
\newcommand{\etcand}{\mbox{$E_{T}^{\subs{cand}}$}}
\newcommand{\etup}{\mbox{$E_{T}^{\subs{up}}$}}
\newcommand{\etdown}{\mbox{$E_{T}^{\subs{down}}$}}
\newcommand{\jet}{\mbox{$E_{T}^{\subs{jet}}$}}
\newcommand{\cet}{\mbox{$E_{T}^{\subs{cell}}$}}
\newcommand{\jetvec}{\mbox{$\vec{E}_{T}^{\subs{jet}}$}}
\newcommand{\cetvec}{\mbox{$\vec{E}_{T}^{\subs{cell}}$}}
\newcommand{\jevec}{\mbox{$\vec{E}^{\subs{jet}}$}}
\newcommand{\cevec}{\mbox{$\vec{E}^{\subs{cell}}$}}
\newcommand{\etfr}{\mbox{$f_{E_{T}}$}}
\newcommand{\aveet}{\mbox{$\langle\et\rangle$}}
\newcommand{\nj}{\mbox{$n_{j}$}}
\newcommand{\ptrel}{\mbox{$p_{T}^{rel}$}}
\newcommand{\etad}{\mbox{$\eta_{d}$}}           %% detector eta
\newcommand{\peta}{\mbox{$\eta$}}               %% physics eta
\newcommand{\aeta}{\mbox{$|\eta|$}}             %% absolute eta
\newcommand{\ifb}{fb$^{-1}$}
\newcommand{\ipb}{pb$^{-1}$}
\newcommand{\inb}{nb$^{-1}$}
\newcommand{\met}{\mbox{${\hbox{$E$\kern-0.63em\lower-.18ex\hbox{/}}}_{T}$}}
\newcommand{\metvec}{\mbox{${\hbox{$\vec{E}$\kern-0.63em\lower-.18ex\hbox{/}}}_{T}\,$}}
\newcommand{\metx}{\mbox{${\hbox{$E$\kern-0.63em\lower-.18ex\hbox{/}}}_{x}\,$}}
\newcommand{\mety}{\mbox{${\hbox{$E$\kern-0.63em\lower-.18ex\hbox{/}}}_{y}\,$}}
\newcommand{\het}{\mbox{$\vec{\mathcal{H}}_{T}$}}
\newcommand{\hetsc}{\mbox{$\mathcal{H}_{T}$}}
\newcommand{\zvrt}{\mbox{$Z$}}
\newcommand{\zcut}{\mbox{$|\zvrt| < 50$}}
\newcommand{\mtwo}{\mbox{$\mathcal{M}_{2}$}}
\newcommand{\mthree}{\mbox{$\mathcal{M}_{3}$}}
\newcommand{\mfour}{\mbox{$\mathcal{M}_{4}$}}
\newcommand{\msix}{\mbox{$\mathcal{M}_{6}$}}
\newcommand{\mn}{\mbox{$\mathcal{M}_{n}$}}
\newcommand{\R}{\mbox{$R_{\subss{MTE}}$}}
\newcommand{\invR}{\mbox{$1/R_{\subss{MTE}}$}}
\newcommand{\eemf}{\mbox{$\varepsilon_{\subss{EMF}}$}}
\newcommand{\echf}{\mbox{$\varepsilon_{\subss{CHF}}$}}
\newcommand{\ehcf}{\mbox{$\varepsilon_{\subss{HCF}}$}}
\newcommand{\eglob}{$\mbox{\varepsilon_{\subs{glob}}$}}
\newcommand{\emte}{\mbox{$\varepsilon_{\subss{MTE}}$}}
\newcommand{\ezvrt}{\mbox{$\varepsilon_{\subss{Z}}$}}
\newcommand{\etot}{\mbox{$\varepsilon_{\subs{tot}}$}}
\newcommand{\Bprime}{\mbox{$\comm{B}^{\prime}$}}
\newcommand{\Nsurv}{\mbox{$N_{\subs{surv}}$}}
\newcommand{\Nfail}{\mbox{$N_{\subs{fail}}$}}
\newcommand{\Ntot}{\mbox{$N_{\subs{tot}}$}}
\newcommand{\p}[1]{\mbox{$p_{#1}$}}
\newcommand{\ep}[1]{\mbox{$\Delta p_{#1}$}}
\newcommand{\delr}{\mbox{$\Delta R$}}
\newcommand{\deleta}{\mbox{$\Delta\eta$}}
\newcommand{\cafone}{{\sc Cafix 5.1}}
\newcommand{\caftwo}{{\sc Cafix 5.2}}
\newcommand{\delphi}{\mbox{$\Delta\varphi$}}
\newcommand{\rphi}{\mbox{$r-\varphi$}}
\newcommand{\etaphi}{\mbox{$\eta-\varphi$}}
\newcommand{\etatphi}{\mbox{$\eta\times\varphi$}}
\newcommand{\Rjet}{\mbox{$R_{jet}$}}
\newcommand{\jphi}{\mbox{$\varphi_{\subs{jet}}$}}
\newcommand{\gphi}{\mbox{$\varphi_{\subs{\gamma}}$}}
\newcommand{\ceta}{\mbox{$\eta^{\subs{cell}}$}}
\newcommand{\cphi}{\mbox{$\phi^{\subs{cell}}$}}
\newcommand{\inlum}{\mbox{$\mathcal{L}$}}
\newcommand{\gm}{\mbox{$\gamma$}}
%\newcommand{\Rjj}{\mbox{\mathcal{R}_{\subs{jj}}}}
% for thesis use this instead:
\newcommand{\Rjj}{\mbox{$\mathbf{R}_{\subs{jj}}$}}
\newcommand{\Rgj}{\mbox{$\mathbf{R}_{\subs{\gamma j}}$}}
\newcommand{\Rmathcal}{\mbox{$\mathcal{R}$}}
\newcommand{\etv}{\mbox{$\vec{E}_{T}$}}
\newcommand{\nvec}{\mbox{$\hat{\vec{n}}$}}
\newcommand{\eprime}{\mbox{$E^{\prime}$}}
\newcommand{\aveprime}{\mbox{$\bar{E}^{\prime}$}}
\newcommand{\geta}{\mbox{$\eta_{\gm}$}}
\newcommand{\jeta}{\mbox{$\eta_{\subs{jet}}$}}
\newcommand{\cjeta}{\mbox{$\eta_{\subs{jet}}^{\subss{CEN}}$}}
\newcommand{\fjeta}{\mbox{$\eta_{\subs{jet}}^{\subss{FOR}}$}}
\newcommand{\cjphi}{\mbox{$\varphi_{\subs{jet}}^{\subss{CEN}}$}}
\newcommand{\fjphi}{\mbox{$\varphi_{\subs{jet}}^{\subss{FOR}}$}}
\newcommand{\etcut}{\mbox{$E_{T}^{\subs{cut}}$}}
%\newcommand{\etg}{\mbox{E_{T\gm}}}
% for thesis, use this instead.
\newcommand{\etg}{\mbox{$E_{T}^{\gm}$}}
\newcommand{\cenet}{\mbox{$E_{T}^{\subss{CEN}}$}}
\newcommand{\foret}{\mbox{$E_{T}^{\subss{FOR}}$}}
\newcommand{\cenen}{\mbox{$E^{\subss{CEN}}$}}
\newcommand{\foren}{\mbox{$E^{\subss{FOR}}$}}
\newcommand{\ejtptc}{\mbox{$E^{\subs{ptcl}}_{\subs{jet}}$}}
\newcommand{\ejtmes}{\mbox{$E^{\subs{meas}}_{\subs{jet}}$}}
\newcommand{\AIDA}{{\sc AIDA}}
\newcommand{\RECO}{{\sc Reco}}
\newcommand{\PYTHIA}{{\sc Pythia}}
\newcommand{\HERWIG}{{\sc Herwig}}
\newcommand{\JETRAD}{{\sc Jetrad}}
\newcommand{\CTone}{\mbox{$|\eta|<0.4}$}
\newcommand{\CTtwo}{\mbox{$0.4\leq|\eta|<0.8$}}
\newcommand{\ICone}{\mbox{$0.8\leq|\eta|<1.2$}}
\newcommand{\ICtwo}{\mbox{$1.2\leq|\eta|<1.6$}}
\newcommand{\FWone}{\mbox{$1.6\leq|\eta|<2.0$}}
\newcommand{\FWtwo}{\mbox{$2.0\leq|\eta|<2.5$}}
\newcommand{\FWthr}{\mbox{$2.5\leq|\eta|<3.0$}}
\newcommand{\LCTone}{\mbox{$|\eta|<0.5$}}
\newcommand{\LCTtwo}{\mbox{$0.5\leq|\eta|<1.0$}}
\newcommand{\LICone}{\mbox{$1.0\leq|\eta|<1.5$}}
\newcommand{\LICtwo}{\mbox{$1.5\leq|\eta|<2.0$}}
\newcommand{\LFWone}{\mbox{$2.0\leq|\eta|<3.0$}}
\newcommand{\CSone}{\mbox{$|\eta|<0.5$}}
\newcommand{\CStwo}{\mbox{$0.5\leq|\eta|<1.0$}}
\newcommand{\CSthr}{\mbox{$1.0\leq|\eta|<1.5$}}
\newcommand{\CSfou}{\mbox{$1.5\leq|\eta|<2.0$}}
\newcommand{\CSfiv}{\mbox{$2.0\leq|\eta|<3.0$}}
\newcommand{\sigA}{\mbox{$\sigma_{\subss{\!A}}$}}
\newcommand{\sigASS}{\mbox{$\sigma_{\subss{A}}^{\subss{SS}}$}}
\newcommand{\sigAOS}{\mbox{$\sigma_{\subss{A}}^{\subss{OS}}$}}
\newcommand{\sigZ}{\mbox{$\sigma_{\subss{Z}}$}}
\newcommand{\sige}{\mbox{$\sigma_{\subss{E}}$}}
\newcommand{\siget}{\mbox{$\sigma_{\subss{\et}}$}}
\newcommand{\sigetone}{\mbox{$\sigma_{\subs{\etone}}$}}
\newcommand{\sigettwo}{\mbox{$\sigma_{\subs{\ettwo}}$}}
\newcommand{\rcal}{\mbox{$R_{\subs{cal}}$}}
\newcommand{\zcal}{\mbox{$Z_{\subs{cal}}$}}
\newcommand{\Runf}{\mbox{$R_{\subs{unf}}$}}
\newcommand{\Rsep}{\mbox{$\mathcal{R}_{sep}$}}
\newcommand{\etal}{{\it et al.}}
\newcommand{\ppbar}{\mbox{$p\overline{p}$}}
\newcommand{\pp}{\mbox{$pp$}}
\newcommand{\qqbar}{\mbox{$q\overline{q}$}}
\newcommand{\ccbar}{\mbox{$c\overline{c}$}}
\newcommand{\bbbar}{\mbox{$b\overline{b}$}}
\newcommand{\ttbar}{\mbox{$t\overline{t}$}}

\newcommand{\bbj}{\mbox{$b\overline{b}j$}}
\newcommand{\bbjj}{\mbox{$b\overline{b}jj$}}
\newcommand{\ccjj}{\mbox{$c\overline{c}jj$}}

\newcommand{\bb}{\mbox{$b\overline{b}j(j)$}}
\newcommand{\cc}{\mbox{$c\overline{c}j(j)$}}

\newcommand{\hboson}{\mbox{$\mathit{h}$}}
\newcommand{\Hboson}{\mbox{$\mathit{H}$}}
\newcommand{\Aboson}{\mbox{$\mathit{A}$}}
\newcommand{\zboson}{\mbox{$\mathit{Z}$}}
\newcommand{\zb}{\mbox{$\mathit{Zb}$}}
\newcommand{\bh}{\mbox{$\mathit{bh}$}}
\newcommand{\btag}{\mbox{$\mathit{b}$}}

\newcommand{\hsm}{\mbox{$h_{SM}$}}
\newcommand{\hmssm}{\mbox{$h_{MSSM}$}}

\newcommand{\prot}{\mbox{$p$}}
\newcommand{\pbar}{\mbox{$\overline{p}$}}
\newcommand{\pt}{\mbox{$p_{T}$}}
\newcommand{\xnot}{\mbox{$X_{0}$}}
\newcommand{\Znot}{\mbox{$Z^{0}$}}
\newcommand{\Wpm}{\mbox{$W^{\pm}$}}
\newcommand{\Wplus}{\mbox{$W^{+}$}}
\newcommand{\Wminus}{\mbox{$W^{-}$}}
\newcommand{\lamb}{\mbox{$\lambda$}}
\newcommand{\nhatbf}{\mbox{$\hat{\mathbf{n}}$}}
\newcommand{\pbf}{\mbox{$\mathbf{p}$}}
\newcommand{\xbf}{\mbox{$\mathbf{x}$}}
\newcommand{\jbf}{\mbox{$\mathbf{j}$}}
\newcommand{\Ebf}{\mbox{$\mathbf{E}$}}
\newcommand{\Bbf}{\mbox{$\mathbf{B}$}}
\newcommand{\Abf}{\mbox{$\mathbf{A}$}}
\newcommand{\Rbf}{\mbox{$\mathbf{R}$}}
\newcommand{\nablabf}{\mbox{$\mathbf{\nabla}$}}
\newcommand{\rarrow}{\mbox{$\rightarrow$}}
\newcommand{\slashp}{\mbox{$\not \! p \,$}}
\newcommand{\slashk}{\mbox{$\not \! k$}}
\newcommand{\slasha}{\mbox{$\not \! a$}}
\newcommand{\slashA}{\mbox{$\! \not \! \! A$}}
\newcommand{\slashpar}{\mbox{$\! \not \! \partial$}}
\newcommand{\intdouble}{\mbox{$\int\!\!\int$}}
\newcommand{\MRSTGU}{MRSTg$\uparrow$}
\newcommand{\MRSTGD}{MRSTg$\downarrow$}
%
% JES:
\newcommand{\Due}{\mbox{$D_{\mathrm{ue}}$}}
\newcommand{\Dth}{\mbox{$D_{\Theta}$}}
\newcommand{\Dof}{\mbox{$D_{\mathrm{O}}$}}
\newcommand{\zbl}{\texttt{ZERO BIAS}}
\newcommand{\mbl}{\texttt{MIN BIAS}}
\newcommand{\mbll}{\texttt{MINIMUM BIAS}}
\newcommand{\nue}{\mbox{$\nu_{e}$}}
\newcommand{\num}{\mbox{$\nu_{\mu}$}}
\newcommand{\nut}{\mbox{$\nu_{\tau}$}}
\newcommand{\mycs}{\mbox{$d^{\,2}\sigma/(d\et d\eta)$}}
\newcommand{\mycsav}{\mbox{$\langle \mycs \rangle$}}
\newcommand{\tdcs}{\mbox{$d^{\,3}\sigma/d\et d\eta_{1} d\eta_{2}$}}
\newcommand{\tdcsav}{\mbox{$\langle d^{\,3}\sigma/d\et d\eta_{1} d\eta_{2} \rangle$}}
\newcommand{\tanb}{$\tan\beta$}
\newcommand{\cotb}{$\cot\beta$}
%%

% Inclusive Definitions
\newcommand{\rstev}{\mbox{$\rs = \T{1.8}$}}
\newcommand{\rssps}{\mbox{$\rs = \T{0.63}$}}
\newcommand{\XX}{\mbox{$\, \times \,$}}
\newcommand{\AP}{\mbox{${\rm \bar{p}}$}}
\newcommand{\SU}{\mbox{$<\! |S|^2 \!>$}}
\newcommand{\ET}{\mbox{$E_{T}$}}
\newcommand{\HT}{\mbox{$S_{{\rm {\sl T}}}$} }
\newcommand{\PT}{\mbox{$p_{t}$}}
\newcommand{\DP}{\mbox{$\Delta\phi$}}
\newcommand{\DR}{\mbox{$\Delta R$}}
\newcommand{\DE}{\mbox{$\Delta\eta$}}
\newcommand{\DEP}{\mbox{$\Delta\eta_{c}$}}
\newcommand{\PH}{\mbox{$\phi$}}
\newcommand{\EA}{\mbox{$\eta$} }
\newcommand{\EAJ}{\mbox{\EA(jet)}}
\newcommand{\AEA}{\mbox{$|\eta|$}}
\newcommand{\Ge}[1]{\mbox{#1 GeV}}
\newcommand{\T}[1]{\mbox{#1 TeV}}
\newcommand{\x}{\cdot}
\newcommand{\ra}{\rightarrow}
\def\D0{D\O}
\def\ETmiss{{\rm {\mbox{$E\kern-0.57em\raise0.19ex\hbox{/}_{T}$}}}}
% units
\newcommand{\mb}{\mbox{mb}}
\newcommand{\nb}{\mbox{nb}}
\newcommand{\rs}{\mbox{$\sqrt{\rm {\sl s}}$}}
\newcommand{\fdel}{\mbox{$f(\DEP)$}}
\newcommand{\fdele}{\mbox{$f(\DEP)^{exp}$}}
\newcommand{\fgap}{\mbox{$f(\DEP\! \geq \!3)$}}
\newcommand{\fgape}{\mbox{$f(\DEP\! \geq \!3)^{exp}$}}
\newcommand{\fpyt}{\mbox{$f(\DEP\!>\!2)$}}
\newcommand{\delth}{\mbox{$\DEP\! \geq \!3$}}
\newcommand{\uplim}{\mbox{$1.1\!\times\!10^{-2}$}}
\def\simge
{\mathrel{\rlap{\raise 0.53ex \hbox{$>$}}{\lower 0.53ex \hbox{$\sim$}}}}
\def\simle
{\mathrel{\rlap{\raise 0.53ex \hbox{$<$}}{\lower 0.53ex \hbox{$\sim$}}}}
% End Inclusive Definitions
%%
\newcommand{\pbarp}{\mbox{$p\bar{p}$}}
\def\ETmiss{\mbox{${\hbox{$E$\kern-0.5em\lower-.1ex\hbox{/}\kern+0.15em}}_{\rm T}$}}
\def\Et{\mbox{$E_{T}$}}
\newcommand{\modeta}{\mid \!\! \eta \!\! \mid}
\def\gevcc{GeV/c$^2$}                   %GeV/c^2
\def\gevc{GeV/c}                        %GeV/c
\def\gev{GeV}                           %GeV
\newcommand{\als}{\mbox{${\alpha_{{\rm s}}}$}}
\def\1960{$\sqrt{s}=1960$ GeV}
\def\etI{E_{T_1}}
\def\etII{E_{T_2}}
\def\itaI{\eta_1}
\def\itaII{\eta_2}
\def\deta{\Delta\eta}
\def\etab{\bar{\eta}}
\def\xq{($x_1$,$x_2$,$Q^2$)}
\def\xx{($x_1$,$x_2$)}
\def\rap{pseudorapidity}
\def\as{\alpha_s}
\def\ap{\alpha_{\rm BFKL}}
\def\apb{\alpha_{{\rm BFKL}_{bin}}}
\def\cm{c.m.}
%%

% The following information is for internal review, please remove them for submission
%\leftline{v2.0991s - to be submitted to PRL}

%\leftline{Andy Haas}

%\rightline{Comment to {\tt d0-run2eb-014@fnal.gov} by May 6, 2009}

\hspace{5.2in}\mbox{FERMILAB-PUB-09-257-E} %the fermi-preprint number

\title{Search for NMSSM Higgs bosons in the {\boldmath $h$\rarrow$aa$\rarrow$\mu\mu~\mu\mu$}, {\boldmath $\mu\mu~\tau\tau$} channels using
{\boldmath \ppbar} collisions at {\boldmath $\sqrt{s}=1.96$}~TeV}

% LIST_OF_AUTHORS_R2.TEX                 5/15/09            
%
\author{V.M.~Abazov$^{37}$}
\author{B.~Abbott$^{75}$}
\author{M.~Abolins$^{65}$}
\author{B.S.~Acharya$^{30}$}
\author{M.~Adams$^{51}$}
\author{T.~Adams$^{49}$}
\author{E.~Aguilo$^{6}$}
\author{M.~Ahsan$^{59}$}
\author{G.D.~Alexeev$^{37}$}
\author{G.~Alkhazov$^{41}$}
% ...

\affiliation{\vspace{0.1 in}(The D\O\ Collaboration)\vspace{0.1 in}}
% ...
\affiliation{$^{6}$University of Alberta, Edmonton, Alberta, Canada;
                Simon Fraser University, Burnaby, British Columbia, Canada;
                York University, Toronto, Ontario, Canada and
                McGill University, Montreal, Quebec, Canada}
% ...
\affiliation{$^{30}$Tata Institute of Fundamental Research, Mumbai, India}
% ...
\affiliation{$^{37}$Joint Institute for Nuclear Research, Dubna, Russia}
% ...
\affiliation{$^{41}$Petersburg Nuclear Physics Institute,
                St. Petersburg, Russia}
% ...
\affiliation{$^{49}$Florida State University, Tallahassee, Florida 32306, USA}
% ...
\affiliation{$^{51}$University of Illinois at Chicago,
                Chicago, Illinois 60607, USA}
% ...
\affiliation{$^{59}$Kansas State University, Manhattan, Kansas 66506, USA}
% ...
\affiliation{$^{65}$Michigan State University,
                East Lansing, Michigan 48824, USA}
% ...
\affiliation{$^{75}$University of Oklahoma, Norman, Oklahoma 73019, USA}
% ...

%\date{\today}
\date{May 19, 2009}

\begin{abstract}
We report on a first search for production of the lightest neutral
CP-even Higgs boson ($h$) in the next-to-minimal supersymmetric
standard model, where $h$ decays to a pair of neutral pseudoscalar
Higgs bosons ($a$), using 4.2~\ifb\ of data recorded with the D0
detector at Fermilab. The $a$ bosons are required to either both
decay to $\mu^+\mu^-$ or one to $\mu^+\mu^-$ and the other to
$\tau^+\tau^-$. No significant signal is observed, and we set limits
on its production as functions of $M_a$ and $M_h$.
\end{abstract}

%Here are the relevant PACS numbers that we can quote in PRL
%(http://www.aip.org/pacs/pacs06/pacs0610.html):

\maketitle

The CERN $e^+e^-$ Collider (LEP) has excluded a SM-like Higgs boson
decaying to \bbbar, $\tau^+\tau^-$ with a mass below 114.4
\gev~\cite{lep}, resulting in fine-tuning being needed in the
minimal supersymmetric SM (MSSM). Slightly richer models, such as
the next-to-MSSM (NMSSM)~\cite{Ellwanger:1996gw}, alleviate this
fine-tuning~\cite{gunion}. The $h$\rarrow\bbbar\ branching ratio
(BR) is greatly reduced because the $h$ dominantly decays to a pair
of lighter neutral pseudoscalar Higgs bosons ($a$). The most general
LEP search yields $M_h>82$~\gev~\cite{opal}, independent of the
Higgs boson decay.

Helicity suppression causes the $a$ boson to decay to the heaviest
pair of particles kinematically allowed. The BR($a$\rarrow$\mu\mu$)
is nearly 100\% for 2$m_\mu$\lt$M_a$$\lesssim$3$m_\pi$ ($\approx$450
MeV) and then decreases with rising $M_a$ due to decay into hadronic
states \cite{Cheung:2008zu}. A M($\mu\mu$) spectrum  in $\Sigma$
decays consistent with $a$\rarrow$\mu\mu$ where $M_a=214.3$~MeV was
reported by the HyperCP collaboration~\cite{hypercp}, which suggests
searching for $h$\rarrow$aa$ with $a$\rarrow$\mu\mu$
\cite{Zhu:2006zv}. Decays to charm are usually suppressed in the
NMSSM, so they have been neglected. If 2$m_\tau$\lt$M_a$\lt2$m_b$,
the BR($a$\rarrow$\mu\mu$) is suppressed by $(M_{\mu}^2/M_{\tau}^2)
/ [\sqrt{1-(2 M_{\tau}/M_a)^2}]$, $a$ decays primarily to
$\tau^+\tau^-$, and the limit from LEP is still weak ($M_h$\gt 86
\gev)~\cite{taus}. The direct search for the 4$\tau$ final state is
challenging, due to the lack of an observable resonance peak and low
$e$, $\mu$ transverse momentum (\pt) which complicates triggering
\cite{Graham:2006tr}. The $2\mu 2\tau$ final state however, contains
a resonance from $a$\rarrow$\mu\mu$, high \pt\ muons for triggering,
and missing transverse energy (\met)~\cite{Lisanti:2009uy}.
B-factories also search for $\Upsilon$\rarrow$a\gamma$, where the
$a$ boson escapes as missing energy or decays to muons or taus
\cite{babar2008st}.

In this Letter, we present a first search for $h$ boson production,
followed by $h$\rarrow$aa$ decay with either both $a$ bosons
decaying to $\mu^+\mu^-$ or one decaying to $\mu^+\mu^-$ and the
other to $\tau^+\tau^-$. Data from Run II of the Fermilab Tevatron
Collider recorded with the D0 detector~\cite{d0det} are used,
corresponding to an integrated luminosity of about 4.2~\ifb. The
signal signature is either two pairs of collinear muons (due to the
low $M_a$), or one pair of collinear muons and either large \met, an
additional (not necessarily isolated) muon, or a loosely-isolated
electron from $a$\rarrow $\tau\tau$ opposite to the muon pair. The
main backgrounds are multijet events containing muons from the decay
of particles in flight ($\pi$, $K$), heavy-flavor decays, and other
sources ($\eta$, $\phi$, $J/\psi$, etc.) and
$Z$/$\gamma^\star$(\rarrow$\mu\mu$)+jets. The {\sc pythia}
\cite{pythia} event generator is used to simulate
$gg$\rarrow$h$\rarrow$aa$ signal events for various $M_h$ and $M_a$,
which are then passed through the {\sc geant}3~\cite{geant} D0
detector simulation and reconstructed.

Events are required to have at least two muons reconstructed in the
muon system and matched to tracks from the inner tracking system
with $p_T>10$~\gev\ and \aeta\lt2, where $\eta$ is the
pseudorapidity. Muons are not required to have opposite electric
charge. No specific trigger requirements are made; an OR of all
implemented triggers is used. But most events selected pass a dimuon
trigger, with muon \pt\ thresholds of 4--6 \gev. Trigger efficiency
is \gt90\% for events passing the offline selections.

For the 4$\mu$ channel, we look for one muon from each of the two
$a$ boson decays, so the dimuon pair with the largest invariant mass
is selected, with $M$($\mu$,$\mu$)\gt15~\gev\ and
$\Delta\cal{R}$($\mu$,$\mu$)\gt1, where
$\Delta\cal{R}=$$\sqrt{(\Delta\eta)^2+(\Delta\phi)^2}$ and $\phi$ is
the azimuthal angle. Only one muon is required to be reconstructed
from each pair of collinear muons. The muon system has insufficient
granularity to reliably reconstruct two close muons. A companion
track is identified with \pt\gt 4 \gev\ and smallest $\Delta\cal{R}$
from each muon, within $\Delta\cal{R}$\lt 1 and $\Delta
z$(track,PV)\lt1~cm, where $z$ is the distance along the beamline,
and PV is the primary \ppbar\ interaction vertex. The muon pair
calorimeter isolation ($\cal I^C_{\mu\mu}$) is the sum of
calorimeter energy within 0.1\lt$\Delta\cal{R}$\lt0.4 of
\emph{either} the muon or the companion track. Both muons are
required to have $\cal I^C_{\mu\mu}$\lt1~\gev\ and track-based
isolation: \lte 3 tracks with \pt\gt 0.5~\gev\ and $\Delta
z$(track,PV)\lt1~cm within $\Delta\cal{R}$\lt0.5 of the muon,
including the muon track itself.

Based on a control data sample greatly enhanced in multijet events
by removing the $\cal I^C_{\mu\mu}$ requirement on the muons, we
predict 1.9$\pm$0.4 events to pass the final selections. The mass of
the leading (trailing) \pt\ muon and its companion track,
$m_1(\mu,{\text{track}})$ ($m_2(\mu,{\text{track}})$), is shown in
the multijet sample in Fig.~\ref{fig:iso_qcd_both_2D}(a) and is used
to model the background shape. Background is also expected from
$Z$/$\gamma^\star$\rarrow$\mu\mu$ events where additional companion
tracks are reconstructed. Studying the dimuon mass distributions in
the isolated data when zero or one of the muons is required to have
a companion track gives an estimate of 0.29$\pm$0.04 events. The
background from \ttbar, diboson, and $W$+jets production is found to
be negligible.

\begin{figure}\centering
\includegraphics[height=1.5in]{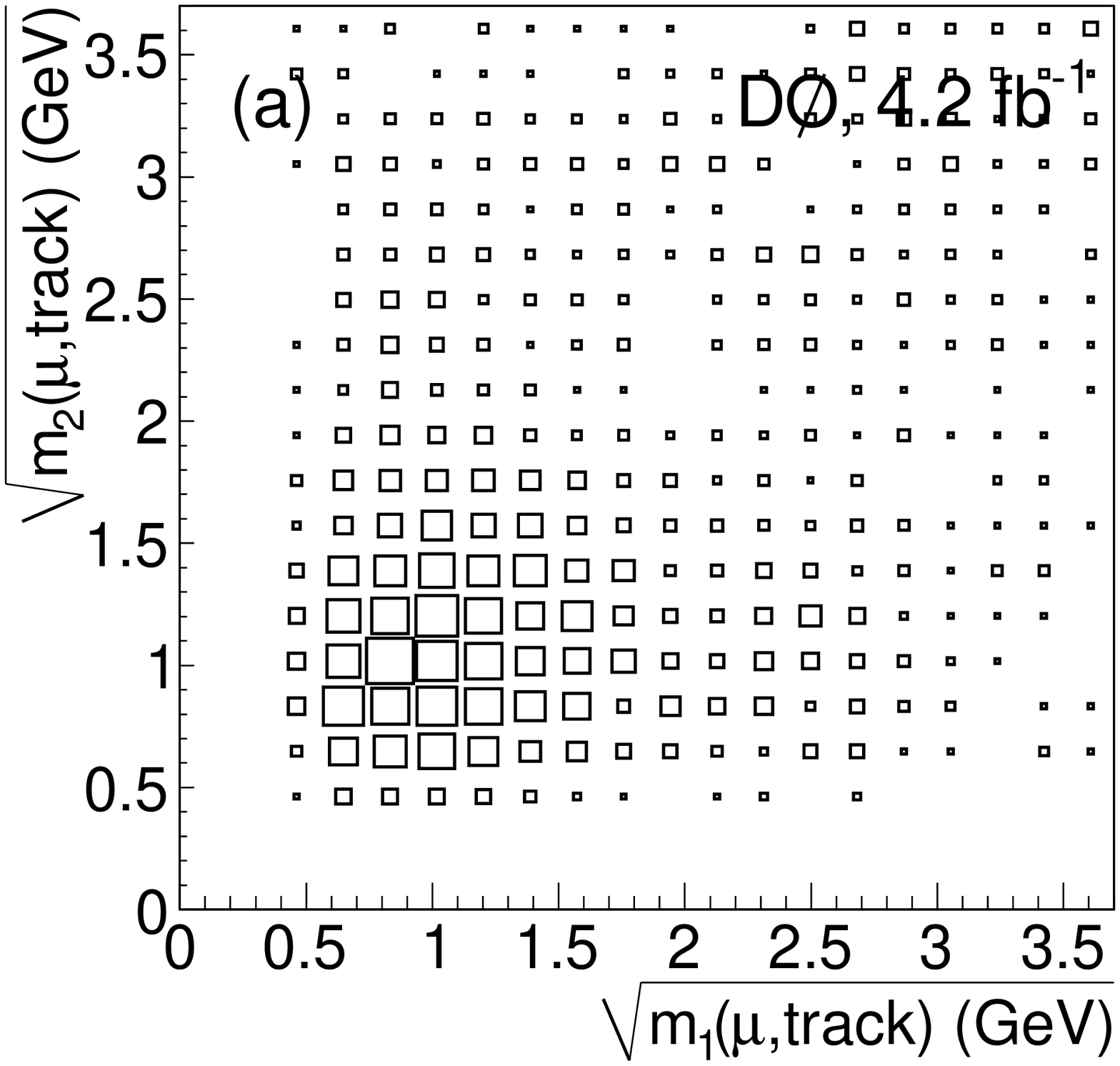}\includegraphics[height=1.5in]{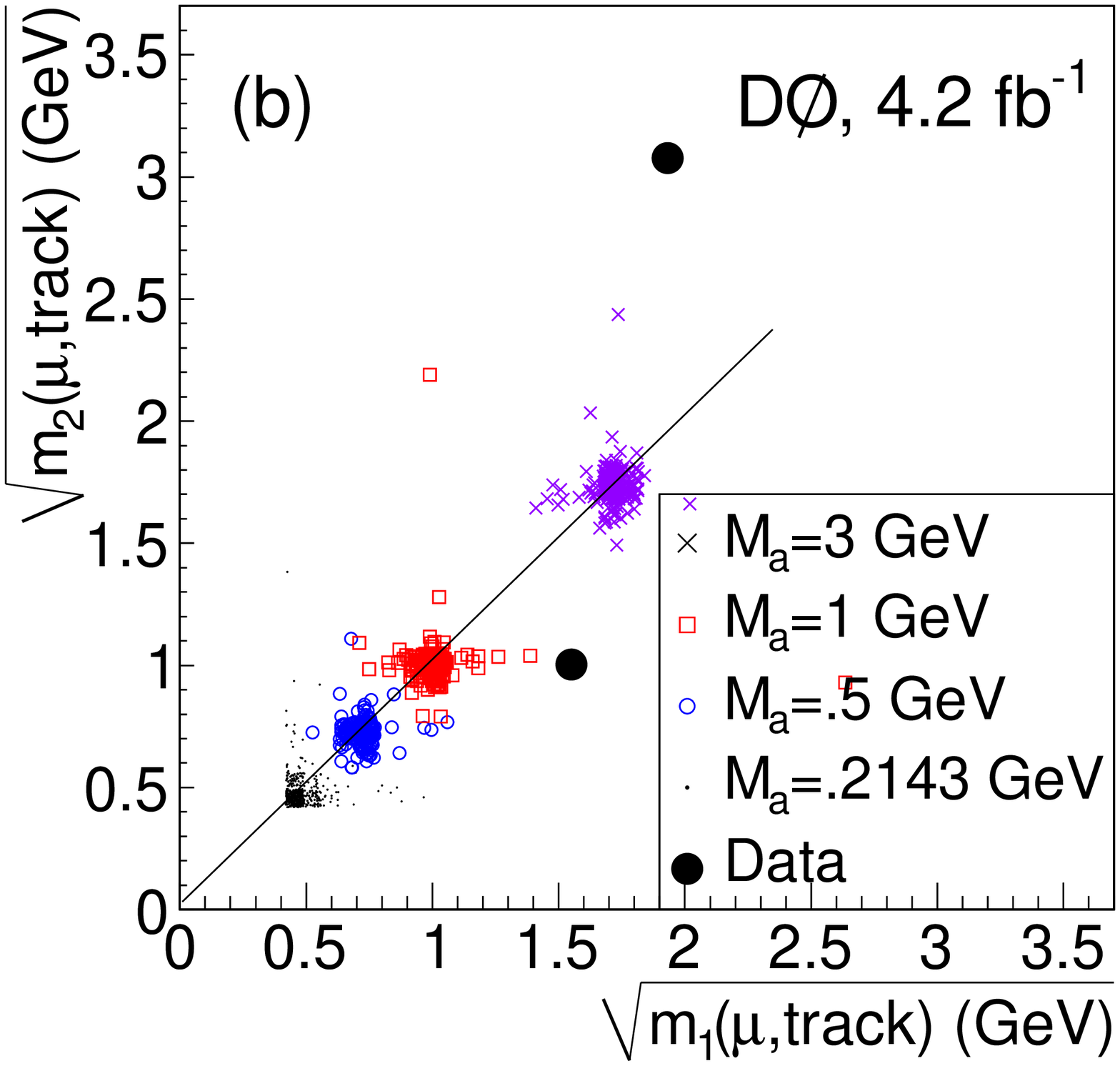}
\caption{The $\sqrt{m_2(\mu,{\text{track}})}$
vs.~$\sqrt{m_1(\mu,{\text{track}})}$ distribution (a) in the
multijet sample, and (b) after the isolation cut is applied to both
muons for data and various MC signal masses.}
\label{fig:iso_qcd_both_2D}
\end{figure}

Signal acceptance uncertainty is dominated by the ability to
simulate the detection of the companion track, particularly when the
two muons are very collinear. We compare $K^0_S$ decays in data and
simulation as a function of the $\Delta\cal{R}$ between the two pion
tracks. Over most of the $\Delta\cal{R}$ range, the relative
tracking efficiency is within 20\%, but few events have
$\Delta\cal{R}$\lt0.02 (corresponding to $M_a$\lt0.5 \gev\ for
$M_h$=100 \gev), and consistency can only be confirmed at the 50\%
level. For $\Delta\cal{R}$($\mu$,$\mu$)\lt 0.1 (corresponding to
$M_a$\lt2 \gev\ for $M_h$=100 \gev), there is the possibility that
the two muons will overlap in the muon system and interfere with
each other's proper reconstruction and triggering. By studying the
effect of adding noise hits, we find up to a 10\% effect on
reconstruction and 20\% effect on the trigger efficiency. The
background uncertainty (50\%) is dominated by the statistical
uncertainty of the multijet-enhanced data sample. The luminosity
uncertainty is 6.1\%~\cite{lumi}.

\begin{table}
\begin{centering}
\caption{\label{tab:4mu_results} The efficiency for MC signal events
within the 2 s.d.~window around each $M_a$, numbers of events
expected from background (with statistical uncertainty) and observed
in data, and the expected and observed limits on the
$\sigma$(\ppbar\rarrow$h$+$X$)$\times$BR($h$\rarrow$aa$\rarrow
4$\mu$), for $M_h$=100 \gev. Limits for other $M_a$, up to
2$m_\tau$, are interpolated from these simulated MC samples. No
events are observed in a window for any interpolated $M_a$.}
\begin{tabular}{cccccc}
  \hline\hline
  $M_a$  & Window & Eff. & $N_{\text{bckg}}$ & $N_{\text{obs}}$ & $\sigma\times$BR \\
   (GeV) & (MeV)  &      &                   &                  & [exp] obs (fb) \\
  \hline
  0.2143 & $\pm$15    & 17\%  & 0.001$\pm$0.001 & 0 & [10.0] 10.0 \\
  0.3    & $\pm$50    & 16\%  & 0.006$\pm$0.002 & 0 & [9.5] 9.5 \\
  0.5    & $\pm$70    & 12\%  & 0.012$\pm$0.004 & 0 & [7.3] 7.3 \\
  1      & $\pm$100   & 13\%  & 0.022$\pm$0.005 & 0 & [6.1] 6.1 \\
  3      & $\pm$230   & 14\%  & 0.005$\pm$0.002 & 0 & [5.6] 5.6 \\
  \hline\hline
\end{tabular}
\end{centering}
\end{table}

After the isolation requirements are applied to both muons, two
events are observed in data, consistent with the total background of
2.2$\pm$0.5 events. Neither has a third muon identified, compared to
about 50\% of the signal MC events. We fit a Gaussian distribution
to the $m_1(\mu,{\text{track}})$ distribution, and the number of
events with \emph{both} $m_1(\mu,{\text{track}})$ and
$m_2(\mu,{\text{track}})$ within a $\pm$2 s.d.~window around the
mean from the fit are determined for data, signal, and background
(Tab.~\ref{tab:4mu_results}). No events are observed within any
window, in agreement with the background prediction. Upper limits on
the $h$\rarrow $aa$\rarrow 4$\mu$ signal rate are computed at 95\%
C.L.~using a Bayesian technique~\cite{limit_calc} and vary slightly
with $M_h$, decreasing by $\approx$10\% when $M_h$ increases from 80
to 150 \gev.

For the 2$\mu$2$\tau$ channel, the muon pair is selected in each
event with the largest scalar sum of muon \pt\ ($\Sigma^{p_T}_\mu$),
with muon \pt\gt 10 \gev, $\Delta\cal{R}$($\mu,\mu$)\lt1, and
$M$($\mu\mu$)\lt20 \gev. This is the ``pre-selection''
(Tab.~\ref{tab:2mu_results}). Next, $\Sigma^{p_T}_\mu$\gt35 \gev\ is
required, to reduce background, and the same muon pair calorimeter
and track isolation cuts are applied as for the 4$\mu$ channel. This
is the ``isolated'' selection.

Standard D0 $\tau$ identification~\cite{Abazov:2008ff} is severely
degraded and complicated by the topology of the two overlapping
$\tau$ leptons. Instead, we require significant \met\ from the
collinear $\tau$ decays to neutrinos. The \met\ is computed from
calorimeter cell energies and corrected for the \pt\ of the muons.
To ensure that this correction is as accurate as possible, the
following additional muon selection criteria are applied. The muons'
tracks in the inner tracker are required to have fits to their hits
with $\chi^2/$dof\lt4, transverse impact parameter from the PV less
than 0.01~cm, and at least three hits in the silicon detector. The
match between the track reconstructed from muon system hits and the
track in the inner tracker must have $\chi^2$\lt 40, and the muon
system track must have \pt\gt 8 \gev. Hits are required for both
muons in all three layers of the muon system. Also, less than
10~\gev\ of calorimeter energy is allowed within
$\Delta\cal{R}$\lt0.1 of either muon, to exclude muons with showers
in the calorimeter. Finally, the leading muon \pt\ must be less than
80 \gev, to remove muons with mismeasured \pt. To improve the \met\
measurement in the calorimeter, the number of jets reconstructed
\cite{RunIIcone} with cone radius $0.5$, \pt\gt15~\gev\ (corrected
for jet energy scale), and \aeta\lt2.5 must be less than five.
Events with \met\gt 80 \gev\ are also rejected to remove rare events
where the \met\ is grossly mismeasured, since signal is not expected
to have such large \met. These are the ``refining'' cuts. Then an
event must pass one of three mutually exclusive subselections. The
first subselection, for when no jet is reconstructed from the tau
pair, requires zero jets with \pt\gt 15~\gev,
$\Delta\phi$($\mu\mu$,\met)\gt2.5, the highest-\pt\ track with
$\Delta z$(track, PV)\lt3~cm and not matching either of the two
selected muon tracks in the dimuon candidate to have \pt\gt 4~\gev\
and $\Delta\phi$(track, \met)\lt0.7. The second subselection, for
when at least one of the tau decays is 1-prong, requires at least
one jet, where the leading-\pt\ jet (jet1) has no more than four
(non-muon) tracks associated with it with \pt\gt0.5~\gev, $\Delta
z$(track,jet1)\lt3~cm, and $\Delta\cal{R}$(track,jet1)\lt0.5,
$\Delta\phi$(jet1,\met)\lt0.7, and \met\gt20~\gev. The third
subselection, for when both tau decays are 3-prong (or more) and
thus most jet-like, requires at least one jet, where jet1 has either
more than four (non-muon) tracks associated with it or
$\Delta\phi$(jet1,\met)\gt0.7 and \met\gt35~\gev. Events passing one
of these three subselections are called the ``\met'' selection.

To gain acceptance, we also select events not passing the ``\met''
selection, but with either an additional muon (not necessarily
isolated) or loosely-isolated electron. For the ``Muon'' selection,
a (third) muon is required, with \pt\gt 4~\gev\ and
$\Delta\phi$($\mu$,\met)\lt0.7. The ``EM'' selection rejects events
in the ``Muon'' selection and then requires an electron with \pt\gt
4~\gev, $\Delta\phi$($e$,\met)\lt0.7, fewer than three jets,
\met\gt10 \gev, and $p_T^e$+\met\gt35~\gev.

The dimuon invariant mass shape of the multijet and $\gamma^\star$
background to the ``\met'' selection is estimated from the low \met\
data which passes the ``refining'' cuts but fails the ``\met''
selection cuts. For the ``Muon'' and ``EM'' selections, it is taken
from the ``isolated'' data sample. The requirements of the ``Muon''
and ``EM'' selections have no significant effect on the dimuon
invariant mass shape for a data sample with loosened isolation
requirements. These background shapes are summed and normalized to
the data passing all selections, but excluding data events within a
2 s.d.~dimuon mass window for each $M_a$ (see below). Background
from diboson, \ttbar, and $W$+jets production, containing true \met\
from neutrinos, is estimated using MC and found to contribute
\lt10\% of the background from multijet and $\gamma^\star$.

\begin{figure}\centering
\includegraphics[height=1.5in]{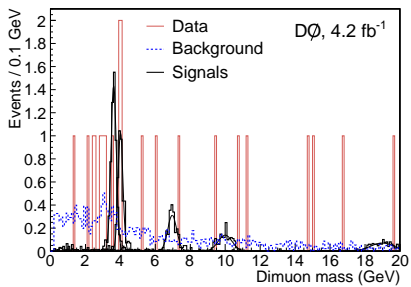}
\caption{The dimuon invariant mass for events passing all selections
in data, background, and 2$\mu$2$\tau$ signals for $M_a=3.6$, 4, 7,
10, and 19 \gev. $\sigma$(\ppbar\rarrow$h$+$X$)=1.9~pb is assumed,
BR($h$\rarrow$aa$)=1, and $M_h$=100 \gev.}
\label{fig:zm_sigback}
\end{figure}

\begin{table*}
\begin{centering}
\caption{\label{tab:2mu_results} Selection efficiencies and limits
for the 2$\mu$2$\tau$ channel, for $M_h$=100 \gev\ and various
$M_a$. The numbers of events at ``pre-selected,'' ``isolated''
stages and after (``refining'') ``\met,'' ``Muon,'' and ``EM''
selections, assuming $\sigma$(\ppbar\rarrow$h$+$X$)=1.9~pb and
BR($h$\rarrow $aa$)=1. Next are the window size, and numbers of
events in the window for signal (and overall efficiency times BR),
expected from background (with statistical uncertainty), and
observed in data. The expected and observed limits on
$\sigma$(\ppbar\rarrow$h$+$X$)$\times$BR($h$\rarrow$aa$) and
$\sigma$(\ppbar\rarrow$h$+$X$)$\times$BR($h$\rarrow$aa$) $\times$ 2
$\times$ BR($a$\rarrow$\mu\mu$)$\times$BR($a$\rarrow $\tau\tau$)
follow.}
%\begin{tiny}
\begin{tabular}{c|c|c|c|c|c|c|c|c|c|c|c}
  \hline\hline
  Sample       & N pre.& N iso.&(ref.) ``\met''&   ``Mu'' &   ``EM'' & Window & $N_{\text{sig}}$ (Eff.) & $N_{\text{bckg}}$ &   $N_{\text{obs}}$ & [exp] obs & $\sigma\times2\times$BR \\
  \hline
  Data         & 95793 & 2795 & (1085) 15  & 4   & 4  &&&&& \\
  $M_a$=3.6 GeV& 53.1  & 28.0 & (14.5) 3.5 & 1.9 &0.8 & $\pm$0.30 GeV& 5.2 (0.066\%) & 1.9$\pm$0.4 & 1 & [1.8] 1.5 pb& [23.8] 19.1 fb\\
  $M_a$=4 GeV  & 33.6  & 15.3 & (8.1)  2.5 & 1.2 &0.4 & $\pm$0.32 GeV& 3.3 (0.042\%) & 1.1$\pm$0.2 & 4 & [2.6] 4.9 pb& [23.9] 45.9 fb\\
  $M_a$=7 GeV  & 20.6  & 8.7  & (4.5)  1.7 & 0.8 &0.3 & $\pm$0.54 GeV& 2.1 (0.027\%) & 1.1$\pm$0.2 & 1 & [4.0] 3.9 pb& [25.0] 24.6 fb\\
  $M_a$=10 GeV & 19.3  & 7.5  & (4.2)  1.1 & 0.6 &0.3 & $\pm$0.95 GeV& 1.5 (0.020\%) & 1.6$\pm$0.3 & 2 & [5.9] 6.5 pb& [24.7] 27.3 fb\\
  $M_a$=19 GeV & 14.6  & 5.4  & (2.9)  0.8 & 0.4 &0.2 & $\pm$1.37 GeV& 1.2 (0.015\%) & 0.6$\pm$0.1 & 1 & [6.3] 7.1 pb& [30.0] 33.7 fb\\
  \hline\hline
\end{tabular}
%\end{tiny}
\end{centering}
\end{table*}

\begin{figure}\centering
\includegraphics[height=1.5in]{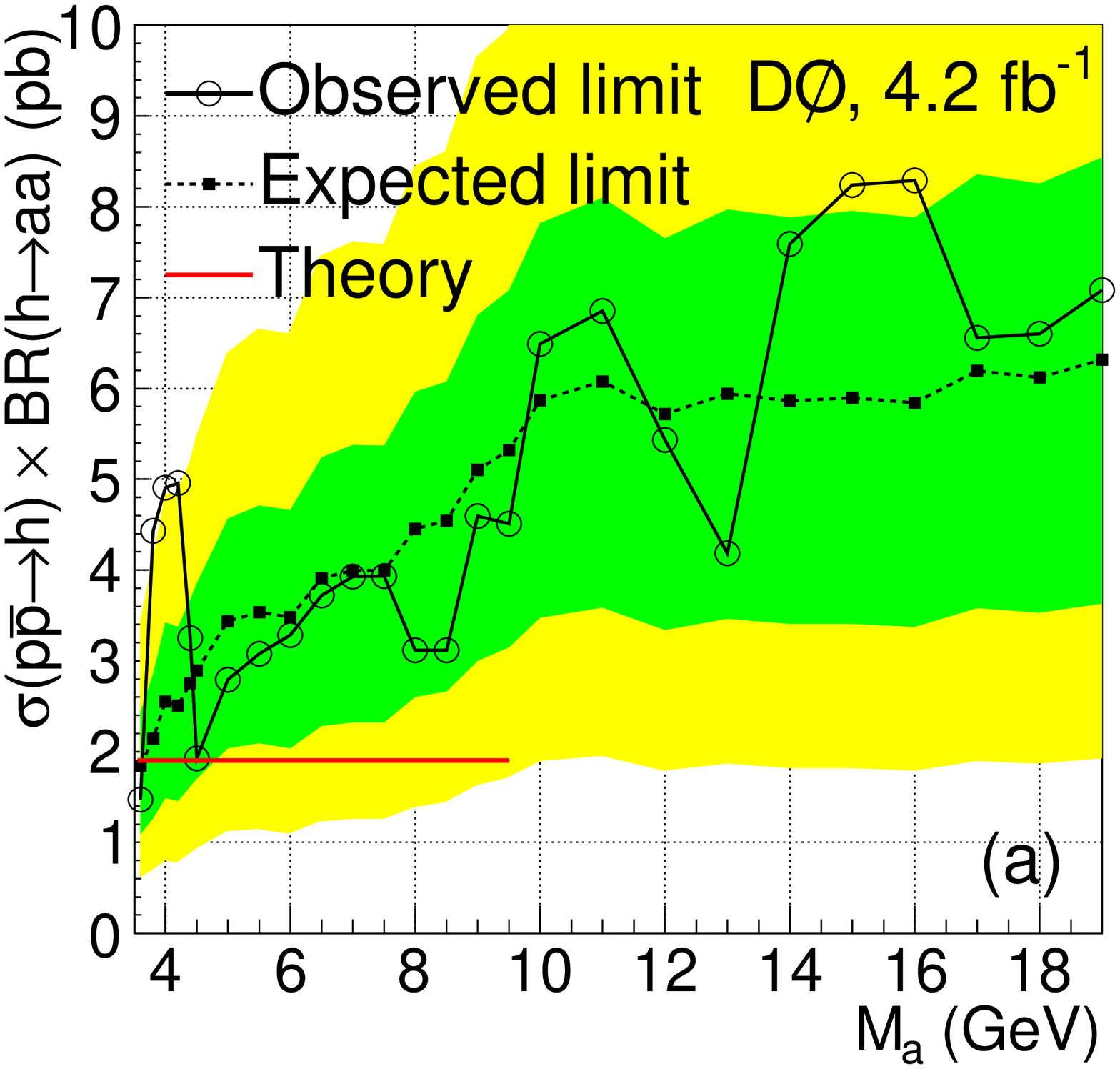}\includegraphics[height=1.5in]{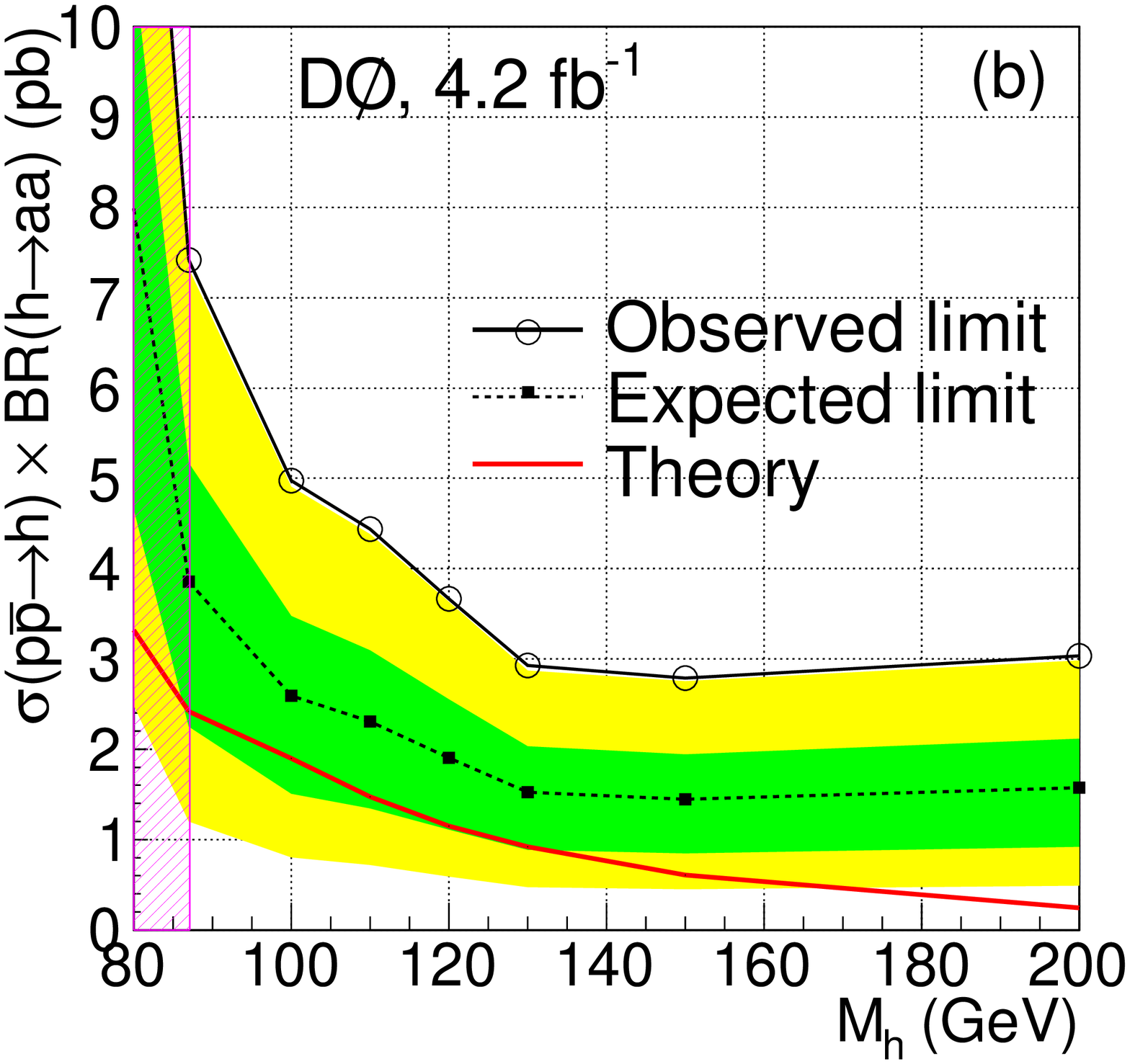}
\caption{The expected and observed limits and $\pm$1 s.d.~and $\pm$2
s.d.~expected limit bands for
$\sigma$(\ppbar\rarrow$h$+$X$)$\times$BR($h$\rarrow$aa$), for (a)
$M_h$=100 \gev\ and (b) $M_a$=4 \gev. The signal for
BR($h$\rarrow$aa$)=1 is shown by the solid line. The region
$M_h$\lt86~\gev\ is excluded by LEP.}
\label{fig:expobs}
\end{figure}

Signal acceptance uncertainty for the 2$\mu$2$\tau$ channel is
dominated by the ability of the simulation to model the efficiency
of the ``refining'' muon cuts and final selections. It is found to
be 20\% per-event based on studies of the muon and event quantities
used, comparing data and MC events in the $Z$ boson mass region.
Comparing the $J/\psi$ and $Z$ boson yields gives a 10\% trigger
efficiency uncertainty. The background uncertainty is less than 20\%
and dominated by the statistical uncertainty of the data sample
used. Alternate fits of the background shape from low \met\ data
modify the background estimates by up to 10\%.

Figure~\ref{fig:zm_sigback} shows the dimuon invariant mass for
data, background, and signals, after all selections. Each signal
dimuon mass peak is fit to a Gaussian distribution, and the numbers
of events with dimuon mass within a $\pm$2 s.d.~window around the
mean from the fit are counted (Tab.~\ref{tab:2mu_results}). Data in
each window are consistent with the predicted background. The
expected and observed limits on the $\sigma\times$BR of the
$h$\rarrow $aa$ process for each $M_a$ studied are shown, assuming
the $a$ boson BRs given by {\sc pythia}, with no charm decays. Since
the $a$ boson BRs are model-dependent, we also derive a result which
factors out the BRs taken from {\sc pythia}. Limits are derived for
intermediate $M_a$ by interpolating the signal efficiencies and
window sizes, see Fig.~\ref{fig:expobs}(a). Above 9.5~\gev, we
expect $a$\rarrow\bbbar\ decays to dominate and greatly decrease
BR($aa$\rarrow2$\mu$2$\tau$), but limits are calculated under the
assumption that the $b$ quark decays are absent. We also study the
limits vs.~$M_h$ for $M_a=4$~\gev, see Fig.~\ref{fig:expobs}(b).

We have presented results of the first search for Higgs boson
production in the NMSSM decaying into $a$ bosons at a high energy
hadron collider, in the 4$\mu$ and 2$\mu$2$\tau$ channels.
%For $M_a$\lt2$m_\tau$, the 4$\mu$ channel is relevant.
The predicted BR($a$\rarrow $\mu\mu$) is driven at low $M_a$ by
competition between decays to $\mu\mu$ and to gluons and has large
theoretical uncertainties~\cite{HiggsHunters}. Therefore, for
$M_a$\lt2$m_\tau$, we set limits only on
$\sigma$(\ppbar\rarrow$h$+$X$)$\times$BR($h$\rarrow
$aa$)$\times$BR$^2$($a$\rarrow $\mu^+\mu^-$), excluding about 10~fb.
Assuming $\sigma$(\ppbar\rarrow$h$+$X$)=1.9~pb~\cite{Hahn:2006my},
corresponding to $M_h$$\approx$100~\gev, BR($a$\rarrow $\mu\mu$)
must therefore be less than 7\% to avoid detection, assuming a large
BR($h$\rarrow$aa$). However, BR($a$\rarrow $\mu\mu$) is expected to
be larger than 10\% for $M_a$\lt 2$m_c$~\cite{Cheung:2008zu}, and
depending on BR($a$\rarrow$c\bar{c}$), which is model-dependent and
typically suppressed in the NMSSM, could remain above 10\% until
$M_a$=2$m_\tau$. Thus these results severely constrain the region
2$m_\mu$\lt$M_a$\lt2$m_\tau$. For $M_a$\gt 2$m_\tau$, the limits set
by the current analysis are a factor of $\approx$1-4 larger than the
expected production cross section.

% acknowledgement_paragraph_r2.tex                         5/15/09
%
We thank the staffs at Fermilab and collaborating institutions, 
and acknowledge support from the 
DOE and NSF (USA);
CEA and CNRS/IN2P3 (France);
FASI, Rosatom and RFBR (Russia);
CNPq, FAPERJ, FAPESP and FUNDUNESP (Brazil);
DAE and DST (India);
Colciencias (Colombia);
CONACyT (Mexico);
KRF and KOSEF (Korea);
CONICET and UBACyT (Argentina);
FOM (The Netherlands);
STFC and the Royal Society (United Kingdom);
MSMT and GACR (Czech Republic);
CRC Program, CFI, NSERC and WestGrid Project (Canada);
BMBF and DFG (Germany);
SFI (Ireland);
The Swedish Research Council (Sweden);
CAS and CNSF (China);
and the
Alexander von Humboldt Foundation (Germany).
%
   % input acknowledgement

\end{document}